\documentclass[a4paper,12pt]{article}

\makeatletter
\renewcommand\paragraph{\@startsection{paragraph}{4}{\z@}%
  {-3.25ex\@plus -1ex \@minus -.2ex}%
  {1.5ex \@plus .2ex}%
  {\normalfont\normalsize\bfseries}}
\makeatother

\usepackage{graphicx}
\usepackage{natbib}
\usepackage{fullpage}
\usepackage{setspace}
\usepackage[noblocks]{authblk}
\usepackage{amssymb}

\def\aj{AJ}
\def\apj{ApJ}

\def\aaps{A\&AS}
\def\mnras{MNRAS}

\def\apss{Ap\&SS}
\def\procspie{Proceedings of the SPIE}
\def\apjs{Astrophysical Journal Supplement Series}

\def\pasj{Publications of the Astronomical Society of Japan}

\def\memras{Memoirs of the Royal Astronomical Society}

\def\qjras{Quarterly Journal of the Royal Astronomical Society}

\usepackage{perpage}
\MakePerPage{footnote}

\title{An Optical-UV-IR Survey of the North Celestial Cap:\\I. The Catalogue}
\author[1]{Evgeny Gorbikov}
\author[2]{Noah Brosch}
\affil[1]{\small{Benoziyo Center for Astrophysics, Faculty of Physics, The Weizmann Institute of Science, Rehovot 76100, Israel}}
\affil[2]{\small{The Wise Observatory and The Raymond and Beverly Sackler School of Physics and Astronomy, The Faculty of Exact Sciences, Tel Aviv University, Tel Aviv 69978, Israel}}

\begin{document}
\onehalfspacing
\renewcommand{\thefootnote}{\fnsymbol{footnote}}
\maketitle

\begin{abstract}

We describe the final product of the North Celestial Cap Survey (NCC Survey, NCCS) - the optical-UV-IR merged catalogue for the region within 10$^{\circ}$ of the North Celestial Pole. The NCC region at $\delta\geq80^\circ$ is poorly covered by modern CCD-based surveys. The optical part of the survey was observed in V, R and I with the Wise Observatory telescopes and was merged with GALEX UV and WISE IR data, producing the catalogue. More than four million objects were observed in at least one optical band. The final catalogue contains $\sim$1.6 million sources observed in all three optical bands, of which some 1.4 million have WISE counterparts and $\sim$300,000 have GALEX counterparts. The astrometric accuracy of the optical NCCS data, derived from a comparison with the UCAC3 catalogue, is better than 0.2 arcsec and the photometry, when compared with SDSS, is good to $\sim$0.15 mag for sources brighter than V = 20.3, R = 21.0 and I = 19.2 mag. The SDSS point-extended source separation is reproduced with 
$>$92\% efficiency.

\end{abstract}

\section{Introduction}

The catalogue described here is the final product of the survey described in \cite{GOR10I}. Originally, the survey was to provide optical support for the Tel Aviv University UV Experiment (TAUVEX; \citealt{ALM07}), an Israeli UV space telescope, by extending the wavelength base of observations from the UV to the optical part of EM spectrum. TAUVEX in its last configuration was intended to observe circumpolar sky regions with $|\delta|>80^\circ$. The North Celestial Cap (NCC) with $\delta\geq+80^\circ$ can be observed throughout the entire year from Israel.

The NCC was poorly covered by modern CCD-based sky surveys. One option of an optical counterpart for the TAUVEX observations in the NCC was the USNO-B1.0 (\citealt{MON03}) catalogue, which fully covers the NCC region. However, it is based on photographic plate scans and suffers from shortcomings of the photographic emulsion; in particular its photometry is not sufficiently accurate. \cite{MON03} examined the USNO-B1.0 photometric data and found systematic errors as high as 0.20 mag and dispersions up to 0.34 mag. The only modern high-precision sky survey available for the NCC region is the Sloan Digital Sky Survey (SDSS; \citealt{SDSS_DR9}), which is sufficiently deep and photometrically accurate. The median photometric errors for point sources at the SDSS limiting magnitudes ($r = 22.2$ and $i = 21.3$ mag) are $\Delta r\cong0.16$ and $\Delta i\cong0.10$ mag. However, SDSS covers only $\sim$12\% of the NCC region.

We originally planned to use the 1-meter telescope of the Wise Observatory with R and I bands for the NCC Survey (NCCS). In this form, as an optical two-colour counterpart for the TAUVEX UV data, the survey was described in \cite{GOR10I}, but it underwent several major changes since then. The TAUVEX mission was canceled in 2010 by the Israeli and Indian Space Agencies while the NCCS optical survey was well underway. However, the NCCS was deemed potentially interesting even lacking the TAUVEX data. The GALEX UV data were used in the project instead of the TAUVEX UV data, although GALEX  did not cover the entire NCC region. Preliminary results and some prospects of the NCCS-GALEX data combination were described in \cite{GOR11}.

Two more changes, not described in \cite{GOR11}, must be mentioned here. First, the R and I data were extended with V-band observations using the C-18 telescope of the Wise Observatory. Second, the WISE All-Sky Data Release became publicly available on March 14, 2012 \footnote{According to the Wide-field Infrared Survey Explorer at IPAC site: \\http://wise2.ipac.caltech.edu/docs/release/allsky/}. We planned to match our optical data to some IR catalogue, e.g., to the Two Micron All Sky Survey (2MASS, \citealt{SKR06}) Point Source Catalogue, but the WISE data are deeper (by $\sim$2 mag for the W1 and W2 bands) and have a longer wavelength coverage in comparison to 2MASS, and therefore, the WISE IR data were added to the NCCS catalogue.

The NCCS catalogue, in its final form presented here, is an \textbf{optical-UV-IR merged catalogue}. It covers gaps in celestial location, wavelength and time between the major modern sky surveys. The visibility of the NCC region from the Northern Hemisphere throughout the entire year allows the NCCS data to be used as a 'pseudo-standard' calibration field for deeper and more extensive sky surveys. The NCC location at intermediate galactic latitudes, about $17^\circ \geq b \geq 37^\circ$, allows the study of both galactic objects, as well as of extragalactic sources. The NCCS data can be used for a general Milky Way (MW) structure study and 3D mapping, as done, e.g.., by \cite{JUR08} using the SDSS data. The NCCS galactic latitude span is potentially interesting for the MW thick disk and transition to halo study, such as done by \cite{BOV12}. The NCCS data might also be helpful for a galactic extinction study using, e.g., the star counting technique, as done by \cite{DOB05}, \cite{GOR10a}, etc. The NCCS wide wavelength coverage, from the UV to the IR, can serve a vast range of tasks, e.g., AGN-candidate colour selection. 

The NCCS comes after the completion of the SDSS photometric observations (DR8 represented the last SDSS photometric expansion) and before the 3$\pi$ Steradian Survey (\citealt{3pi_survey}) of the Panoramic Survey Telescope \& Rapid Response System (Pan-STARRS\footnote{http://pan-starrs.ifa.hawaii.edu/public/home.html}) project data become publicly available. 

Some scientific implications of the NCCS catalogue will be described in subsequent papers. Here we concentrate on technical aspects of the survey - the data acquisition and reduction, the catalogue compilation and its merging with other data sets, and its validation with SDSS data.

\section{Data Acquisition and Reduction}
\label{sec:data_ar}

\subsection{Optical Observations}

The procedures of data acquisition and reduction of the optical part of our survey were tested on a $\sim$14 deg$^2$ sky region and were described in \cite{GOR10I}. The major part of the final pipeline used here is similar to that described in there, with only minor changes. Below, the \citeauthor{GOR10I} description is followed, with emphasis on the changes.

\cite{COU76} R and I images were collected with the 1-meter Ritchey-Chr\'{e}tien telescope and the Large Area Imager for the Wise Observatory (LAIWO, \citealt{GOR10I}). LAIWO is a four-CCD mosaic camera imaging a non-contiguous area of $\sim$0.98$\times$0.98 deg$^2$ at $\sim$0.87 arcsec per binned pixel. Each CCD uses four read-out channels, producing 16 image quadrants.

\cite{JOH63} V-band images were obtained with the 0.46-meter Centurion 18 (C18) prime-focus telescope of the Wise Observatory (\citealt{BRO08}). The C18 camera uses a single CCD imaging a contiguous field of view of $\sim$1.25$\times$0.83 deg$^2$ at $\sim$1.47 arcsec per pixel.

R and I data were collected on 83 nights, from January 10, 2009 to May 2, 2011. Each night several fields were observed with three 300-sec exposures in R and the same in I, with small $\sim$10 arcsec dithering. The V data were collected on 102 nights from June 24, 2011 to May 4, 2012. To match the R- and I-band survey depth, six V-band 600-sec exposures with $\sim$5 arcsec dithering were obtained for each field. We preferentially observed fields with a small ($\sim$15 arcmin on average) overlap, located at the lowest possible airmass ($<2$, preferably), aiming also to cover the gaps between the LAIWO CCDs to ensure 100\% sky coverage in all three optical bands in the survey area. The typical seeing at the Wise Observatory is $\sim$2--3 arcsec\footnote{See \cite{BRO92} for a more detailed description of the observing conditions at the Wise Observatory.}. The seeing was better than 1.5 arcsec during $\sim$8\% of the nights. When the seeing was worse than 3.5 arcsec, the observations were not used for this study.  

The data reduction used an automated pipeline written mostly in IRAF (Image Reduction and Analysis Facility; \citealt{TOD86}), with some integrated modules written in other programming software. The pipeline processes an entire night directory of raw images at once, extracting objects and adding them to the catalogue.

The LAIWO electrical setup (see \citealt{GOR10I}) produces cross-talk between different read-out channels of the same CCD, resulting in the appearance of ghost images. The phenomenon is produced for every pixel, but is evident for saturated pixels. Ghost images can be easily confused with real ones or can even merge with the image of a real object, and must be removed by de-ghosting.

The cross-talk coefficients between the different read-out channels are either calculated using images from the same night, or are derived from cross-talk coefficients of neighbouring nights using interpolation. The coefficients are $\sim$0.001-0.0001, are different for different quadrants, but are approximately similar for the same quadrant in different nights. They do not depend on the filter and do not demonstrate long-period variations. The ghost images are removed by subtracting each quadrant multiplied by its cross-talk coefficient from the other quadrants.

The de-ghosted images are debiased, dark-subtracted and flat-field-corrected. Three 300-sec LAIWO images of the same field taken with the same filter, or six 600-sec C18 images, are registered and median-combined. Assuming that the sky brightness did not change significantly between the exposures, this improves the signal-to-noise (S/N) ratio of the final image and removes random noise, such as dust patterns, cosmic rays, satellite tracks, etc.

The combined science images are then scanned with SourceExtractor (SE, \citealt{BER96}), which was set to detect all the sources 2$\sigma$ above the noise. This produces a file with a list of different parameters for all detected sources. The photometry used to extract sources is \cite{KRO80} photometry using an elliptic adaptive aperture, the natural choice for the NCCS (\citealt{GOR10I}).

The pixel coordinates of all objects were transformed to J2000.0 $(\alpha, \delta)$ using UCAC3 (The Third USNO CCD Astrograph Catalogue, \citealt{ZAC10}) objects in the neighbourhood of the field. The major advantages of using UCAC3 are all-sky coverage and high astrometric precision (15 to 100 mas; \citealt{ZAC10}). However, UCAC3 contains only relatively bright (R $\lesssim$ 16.3 mag) objects, and, since LAIWO images saturate at R $\cong$ 11.5 mag, only a relatively small overlap of $\sim$five magnitudes is available between the two surveys.

The UCAC3 coordinates were transformed to (X ,Y) planar coordinates using the middle point of the LAIWO array for the R and I images, or the middle point of C18 CCD for the V images as the tangential point. The (X, Y) list of the detected sources is matched to the UCAC3 (X, Y) list by minimizing the root-mean-square (RMS) distance between $\sim$250--350 unsaturated sources per CCD using the downhill simplex algorithm (\citealt{NEL65}). The final astrometric solution was obtained using a tangential projection, allowing for distortion along the X and Y axes. The RMS distance between the matched NCCS points and UCAC3 was usually $\lesssim$0.2 arcsec in each coordinate, thus 0.2 arcsec is defined to be the NCCS astrometric accuracy in both $\alpha$ and $\delta$.

The pipeline also normalizes the object Full-Width-at-Half-Maximum (FWHM) by the mode FWHM value of the field. This relative FWHM value, denoted as $\epsilon$FWHM, is required for the point/extended source separation. The pipeline then joins all the SE output files from the same night into a night catalogue.

The object magnitudes in the night catalogues are calibrated photometrically using \cite{LAN09} standard stars observed together with regular survey fields during several nights, or with secondary standards. The nights with \cite{LAN09} standard stars were included first in the catalogue.

The following calibrations were derived:
\begin{equation}
\begin{array}{rcl}
 V,R,I&=&-2.5\log\frac{N_{i}}{t}-k_{i}X+Z_{i}+C_{i}(R-I),\\
\end{array}
\label{eq:calibVRI}
\end{equation}
where $i$ denotes the filter V, R or I, $\frac{N}{t}$ is the object flux in ADU per sec, $X$ is the object airmass, $k$ is the atmospheric extinction coefficient, $Z$ is the zero point and $C$ is the colour-correction term. The coefficients $k$, $Z$ and $C$ are wavelength-dependent and are determined from the observed \cite{LAN09} standard stars. The atmospheric extinction coefficient and the zero point depend on observation conditions and may vary from night to night. The colour-correction term depends primarily on the instrument, particularly on the filter characteristics, and should be constant on all the nights.

The colour-correction terms $C_i$ were estimated using \citeauthor{LAN09} standard stars for the nights when those standards were observed, and remained constant within the error range from night to night. Their mean values are shown in Table \ref{tab:cc_terms}; the small values for the V and R band imply that the filters used here are similar to those used by \cite{LAN09}.

\begin{table}[hb!]
\caption{Mean Color-Correction Terms and Their Errors.\label{tab:cc_terms}}
\begin{center}
 \begin{tabular}{|c||c|c|c|}
\cline{2-4}
\multicolumn{1}{c||}{} & \multicolumn{3}{c|}{Filter} \\
\cline{2-4}
\multicolumn{1}{c||}{} & V & R & I \\
\hline\hline
C & -0.0421  & -0.0002  & 0.2198  \\
\hline
$\Delta$C & 0.0059 & 0.0049 & 0.0087 \\
\hline
 \end{tabular}
\end{center}
\end{table}

Equation (\ref{eq:calibVRI}) without the last term defines magnitudes $m_V$, $m_R$ and $m_I$, that are not colour-corrected (uncorrected magnitudes):
\begin{equation}
\begin{array}{rcl}
 m_i&=&-2.5\log\frac{N_i}{t}-k_iX+Z_i.\\
\end{array}
\label{eq:calibmVmRmI}
\end{equation}

For nights when \cite{LAN09} standards were observed, the coefficients $k$ and $Z$ were derived directly from the standards and the uncorrected magnitudes $m_V$, $m_R$ and $m_I$ were calculated immediately and saved in the night catalogue. For nights when no photometric standards were observed the calibration was derived by using secondary standards, objects located in the overlap regions between fields that were calibrated in one of the photometric sessions. Assuming that the airmass and the atmospheric extinction were similar for all the objects in the same field, the difference between the uncorrected magnitudes of these objects from different nights will be constant:
\begin{equation}
\begin{array}{rcl}
m_i&=&-2.5\log\frac{N_i}{t}+ZP_i.\\
\end{array}
\label{eq:calib2}
\end{equation}
Here $ZP$ are new zero points, which include the old zero points and the extinction-correction term and are, therefore, constant for all the objects in the same field. The zero points $ZP$ were derived from the secondary standards of a particular field, and the uncorrected magnitudes were calculated for all the objects of that field and stored in the night catalogue. In this manner the photometric calibration was ``pulled'' from field to field, and from calibrated nights to nights when no standard stars were observed.

The full calibrations in Equation (\ref{eq:calibVRI}) cannot be applied directly to the object fluxes, since the colour $(R-I)$ is not known for each object and, therefore, the colour correction cannot be performed. However, the $(R-I)$ colour can be derived from Equation (\ref{eq:calibVRI}) using the uncorrected magnitudes and simple algebra:
\begin{equation}
 R-I=\frac{m_R-m_I}{1-(C_R-C_I)}.
\label{eq:RIdef}
\end{equation}
Hence, to allow the $(R-I)$ colour derivation, a source should have both $m_R$ and $m_I$ magnitudes, and the entries in different bands associated with the same source must be first matched. The derived $(R-I)$ colours can then be used in the full Equation (\ref{eq:calibVRI}) to complete the colour correction. The colour-corrected magnitudes V, R and I are stored only in the final catalogue. 

\begin{table}[ht!]
\caption{Explanations to the columns of the optical catalogue stored in file \textit{catalog\_final.mat}. \label{tab:catalog_final}}
\begin{center}
 \begin{tabular}{|c|l|c|}
 \hline
\textbf{Column} & \textbf{Explanation} & \textbf{Units}\\
\hline\hline
(1) & Catalogue object name. & \\\hline
(2) & Object status. & \\\hline
(3), (5) & J2000.0 $(\alpha, \delta)$ coordinates of objects. & [deg]\\\hline
(4), (6) & Coordinate errors, $(\Delta\alpha, \Delta\delta)$. &[arcsec] \\\hline
(7), (8) & Object V magnitude and its error. & [mag] \\\hline
(9), (10) & Object R magnitude and its error. & [mag] \\\hline
(11), (12) & Object I magnitude and its error. & [mag] \\\hline
(13), (14), (15) & Object relative FWHM, $\epsilon$FWHM. & \textit{unitless}\\
& V, R and I band respectively. & \\\hline
(16), (17), (18) & SE internal flags. & \textit{unitless}\\
& V, R and I band respectively. & \\\hline
\end{tabular}
\end{center}
\end{table}

As shown in Table \ref{tab:VRIcount}, a large fraction of unique sources in the NCCS were observed more than once, and thus have several entries in night catalogues. These entries were matched based on distance, using a matching radius of 3 arcsec ($\sim$2 C18 pixels), and were entered as a single source in the final optical catalogue. The objects that were observed more than once are located in the overlapping fields. We derived RMS magnitude differences for these objects. The RMS differences were found to be less than 0.014 mag for the V band, 0.017 mag for the R band and 0.020 mag for the I band. A few objects observed at $\delta<+80^\circ$ are also included in the final catalogue.

\begin{table}[ht!]
\caption{Number of NCCS sources observed in each optical band: total number, number and fraction of sources observed more than once ($>1\times$), more than twice ($>2\times$) and more than three times ($>3\times$). \label{tab:VRIcount}}
\begin{center}
 \begin{tabular}{|c||c|c|c|c|}
 \hline
\textbf{Band} & \multicolumn{4}{c|}{\textbf{Number of Sourses Observed}}\\
\cline{2-5}
 & \textbf{Total} & $>1\times$ & $>2\times$ & $>3\times$ \\
\hline\hline
 V & 1,939,920 & 690,610 (35.6\%) & 179,200 (9.2\%) & 40,081 (2.1\%) \\
 R & 2,959,975 & 1,392,277 (47.0\%) & 601,613 (20.3\%) & 197,635 (6.7\%) \\
 I & 3,142,200 & 1,569,801 (50.0\%) & 703,675 (22.4\%) & 237,817 (7.6\%) \\
\hline
\end{tabular}
\end{center}
\end{table}

The final NCCS optical catalogue was stored as a single MATLAB file named \textit{catalog\_final.mat}. The file contains an array of 18 columns and the example of the array data is shown in Appendix A. Table \ref{tab:catalog_final} gives a brief description of the array column headers. 

The object name in final catalogue is a number counted in order of object inclusion. The object status can be 1, 2 or 3. Status 3 implies that the object was observed in all three optical filters, thus its V, R and I magnitudes are calibrated and colour-corrected. Status 2 means that the object was observed in R and I only and lacks a V counterpart; the R and I magnitudes are calibrated and colour-corrected while V is missing. Status 1 means that the object was not observed in either R or I or in both bands, thus all its magnitudes are not calibrated nor are colour-corrected.

The object coordinates are calculated as the weighted means\footnote{For a sample of $x_i$ measurements, each measured with an error $\sigma_i$, the weighted mean is defined as $\overline{x}=\frac{\sum x_i/\sigma^2_i}{\sum 1/\sigma^2_i}$, and the error of weighted mean is $\sigma(\overline{x})=\frac{1}{\sum 1/\sigma^2_i}$ (see \citealt{BAR89}, Chapter 4.2.2).\label{page:weighted_mean}} of the coordinates and magnitudes of all the entries associated with the object. The errors are calculated as the error of the weighted mean. For the mean magnitudes and their errors we use only the entries of the corresponding filter. The mean magnitudes are then colour-corrected for objects with status 2 and 3, and their error is increased by the corresponding value.

The object relative FWHM, $\epsilon$FWHM, is calculated as a simple arithmetic mean of $\epsilon$FWHMs of the corresponding entries for each filter. The SE internal flag value for each object in each filter is taken as a maximal value of the SE flag for all corresponding entries, i.e. an object which is 'clean' (internal SE flag = 0) in the final catalogue was 'clean' in every night catalogue whenever it was observed. The list of the SE internal flags is shown in the SExtractor Manual V2.5\footnote{ftp://ftp.iap.fr/pub/from\_users/bertin/sextractor/}.

The final NCCS optical catalogue includes all the entries from all the nights and contains 4,102,998 unique objects, but not all were observed in all three optical bands. About 44.8\% of these objects were observed in one filter only and can be false detections. About 16.8\% of the objects were observed in both R and I filters but not in V, and are mostly faint. Only 1,569,292 objects ($\sim$38.3\% of the unique entries in the final catalogue) were detected in all three optical filters.

An example of the NCCS optical data colour-colour and colour-magnitude diagrams is shown in Appendix B.

\subsection{Catalogue Cleaning, Adding PESS, WISE, GALEX}

\begin{figure*}[ht!]
 \begin{center}
\includegraphics[angle=0,width=0.6\textwidth]{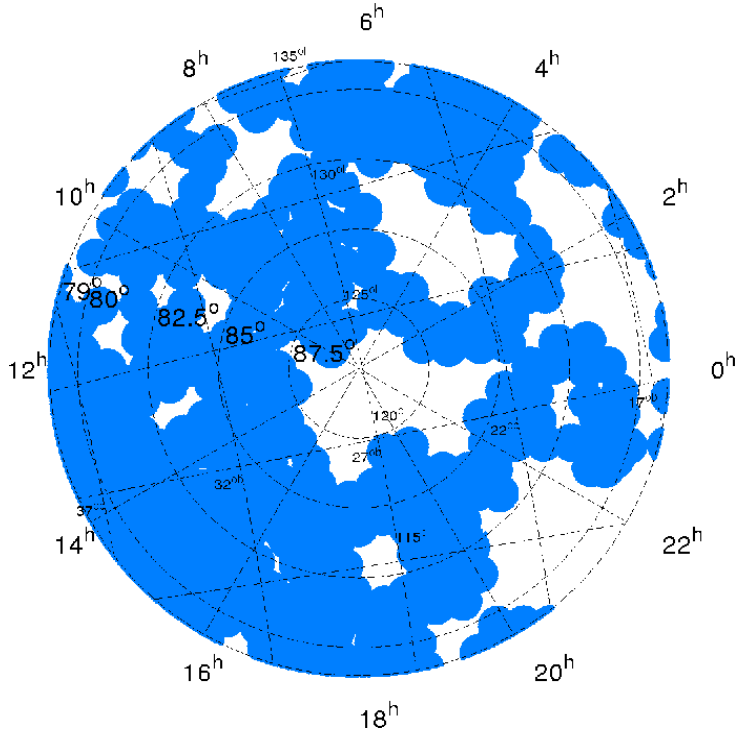} 
\end{center}
 \caption{GALEX coverage of the region $\delta>79^\circ$.
    \label{Fig:zones}}
\end{figure*}

Some further procedures require data in all three optical bands; e.g., the point-extended-source-separation (PESS) routine, which is different from that in our previous paper (\citealt{GOR10I}). Therefore the final catalogue was cleaned of all the objects with status $\neq$ 3, i.e., only 1,569,292 objects observed in all three optical bands were retained.

The PESS procedure adds a value named 'PESS score' to each object in the cleaned catalogue. The PESS procedure is described below in Section \ref{ch:pess}. A PESS score of 0 or 1 means a point-like object; a value of 2 or 3 is for an extended object. 

\begin{figure*}[ht!]
 \begin{center}
\includegraphics[angle=0,width=0.97\textwidth]{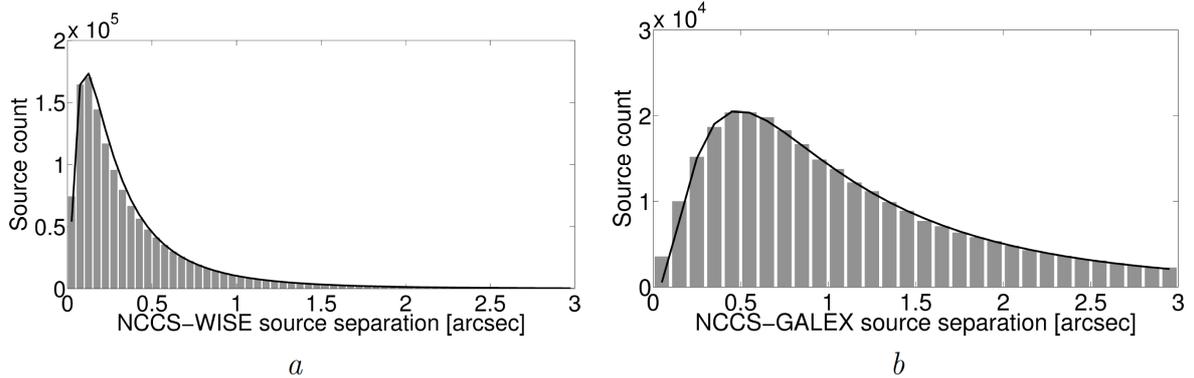}
\end{center}
 \caption{Source count as a function of the great-circle distance between the matched NCCS and WISE ($a$) sources and the matched NCCS and GALEX ($b$) sources. Black solid lines are the fitted log-normal distributions.
    \label{Fig:NCCS_distances}}
 \end{figure*}

The cleaned catalogue was matched with sources from the Wide-field Infrared Survey Explorer (WISE; \citealt{WRI10}) All-Sky
Release data, downloaded from the Infrared Science Archive site\footnote{http://irsa.ipac.caltech.edu/Missions/wise.html}. The WISE astrometry is better than 0.15 arcsec ``\textit{for high S/N sources}'' (\citealt{WRI10}), comparable with the NCCS optical precision (see Section \ref{sec:astrometry}). We used the same matching radius of 3 arcsec for the optical-WISE data combination. The total number of matched WISE objects is 1,357,281, $\sim$86.5\% of the clean catalogue. Fig.~\ref{Fig:NCCS_distances}$a$ shows the distribution of the separation between the matched NCCS and WISE sources fitted with a log-normal model with an average distance of 0.11$^{+0.17}_{-0.06}$ arcsec ($\chi^2/DoF = 58.8/24$).

We also matched between NCCS sources and Galaxy Evolution Explorer (GALEX; \citealt{MOR07}) data. All the GALEX data available for the survey region are derived from the All Sky Survey (AIS) GR6 data release and the data were downloaded from the Mikulski Archive for Space Telescopes site\footnote{http://galex.stsci.edu/GR6/}. The GALEX positional uncertainty is $\sim$0.5 arcsec, greater than the NCCS optical astrometric precision, but less than the matching radius used before. Therefore, the same 3 arcsec radius was used to match between the optical and the GALEX data. The GALEX coverage in the survey region is only $\sim$50\% and not contiguous, as shown in Fig.~\ref{Fig:zones}. The number of FUV objects is one order of magnitude less than the number of NUV objects. GALEX counterparts were found for 301,482 optical sources, $\sim$19.2\% of the clean catalogue sources. Fig.~\ref{Fig:NCCS_distances}$b$ shows the distribution of the separation between the matched NCCS and GALEX sources fitted with a log-normal model, with an average distance of 0.49$^{+0.64}_{-0.28}$ arcsec ($\chi^2/DoF=67.0/25$).

The final merged catalogue was stored in a single MATLAB file \\ \textit{catalog\_clean\_wise\_galex.mat}, which contains an array of  39 columns, described in Table \ref{tab:catalog_cwg}. An example of the merged catalogue data is shown in Appendix C.

\begin{table}[ht!]
\caption{Explanations to the columns of the final merged catalogue stored in file \textit{catalog\_clean\_wise\_galex.mat}. \label{tab:catalog_cwg}}
\begin{center}
 \begin{tabular}{|c|l|c|}
 \hline
\textbf{Column} & \textbf{Explanation} & \textbf{Units}\\
\hline\hline
(1) - (18) & As in Table \ref{tab:catalog_final}. & \\\hline
(19) & PESS score (see Section \ref{ch:pess}). & \\\hline
(20), (21) & J2000.0 $(\alpha, \delta)$ coordinates of matched WISE source. & [deg]\\\hline
(22), (23) & WISE coordinate errors, $(\Delta\alpha, \Delta\delta)$. &[arcsec] \\\hline
(24) & Cross-term of WISE coordinate errors, $\sigma_{\alpha\delta}$. & [arcsec] \\\hline
(25), (26) & WISE source W1 magnitude and its error. & [mag] \\\hline
(27), (28) & WISE source W2 magnitude and its error. & [mag] \\\hline
(29), (30) & WISE source W3 magnitude and its error. & [mag] \\\hline
(31), (32) & WISE source W4 magnitude and its error. & [mag] \\\hline
(33), (34) & J2000.0 $(\alpha, \delta)$ coordinates of matched GALEX source. & [deg] \\\hline
(35), (36) & GALEX source NUV magnitude and its error. & [mag] \\\hline
(37), (38) & GALEX source FUV magnitude and its error. & [mag] \\\hline
(39) & AGN candidate flag. & \\\hline
\end{tabular}
\end{center}
\end{table}

\subsection{Public Access to the Catalogue}

The NCCS catalogue is available for download in two versions from the Wise observatory ftp server and can be used publicly. To ease the catalogue access, it was saved in ASCII coma-separated format. The full optical uncleaned NCCS catalogue, containing 4,102,998 objects, is available at:\\
ftp://wise-gate.tau.ac.il/pub/evgenyg/NCCS\_catalog/catalog\_final.dat \\
The cleaned and matched optical-UV-IR catalogue, containing 1,569,292 objects, is available at:\\ 
ftp://wise-gate.tau.ac.il/pub/evgenyg/NCCS\_catalog/catalog\_clean\_wise\_galex.dat 

\section{Data Characterization and Comparison with SDSS}
\label{sec:SDSScomp}

\subsection{NCCS Depth}

\begin{figure*}[ht!]
 \begin{center}
 \includegraphics[angle=0,width=0.97\textwidth]{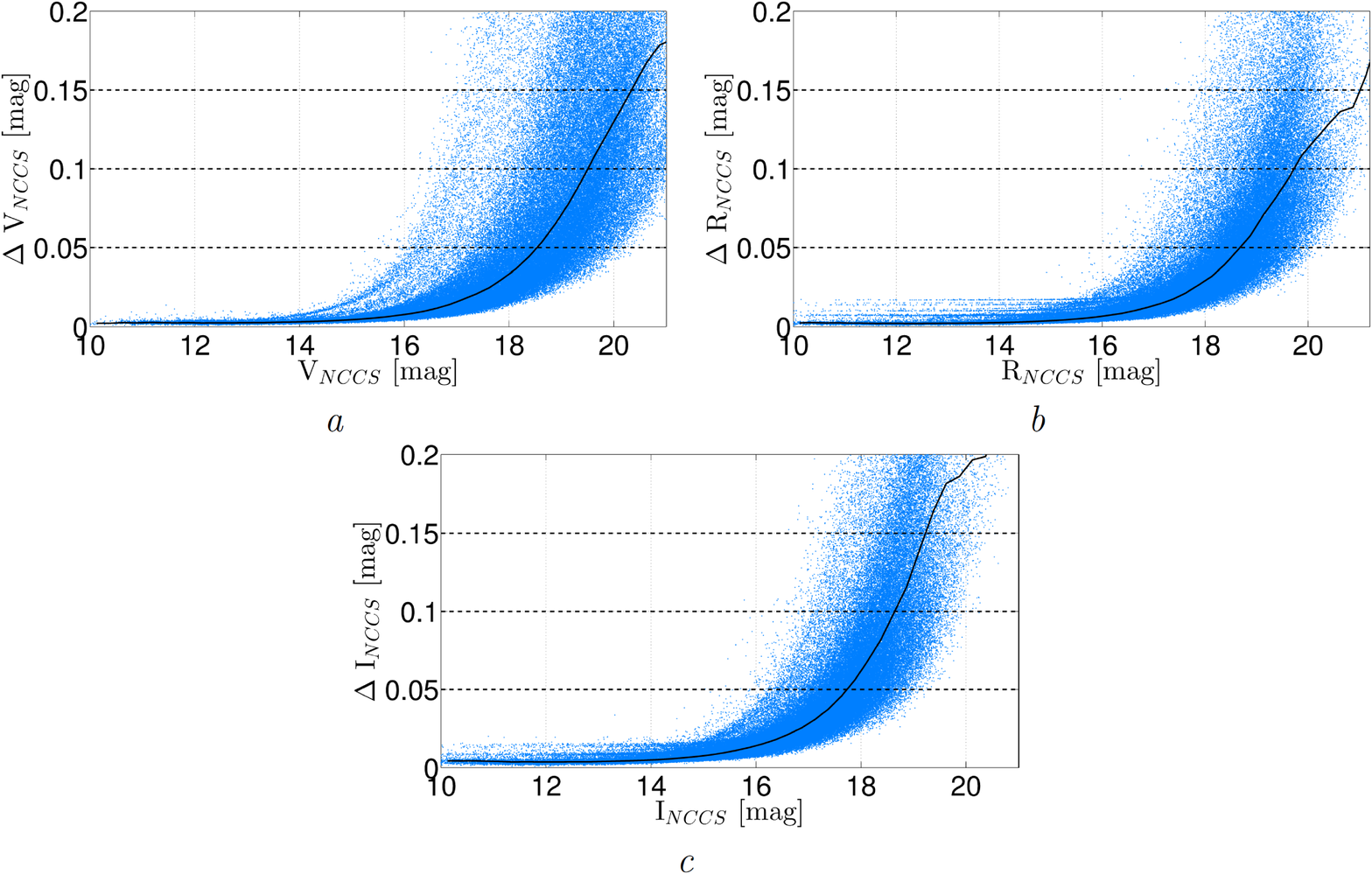} 
 \end{center}
 \caption{Distribution of photometric errors as a function of magnitudes of the NCCS sources. Panels \textit{a}, \textit{b} and \textit{c} show the data for the V, R and I band respectively. The black solid line shows the running median of photometric errors. The black dashed lines represent the limits of 0.05, 0.1 and 0.15 mags.
    \label{Fig:lim_mag}}
 \end{figure*}

We tested the NCCS depth and the results are shown in Fig.~\ref{Fig:lim_mag}. Table \ref{Tab:lim_mag} shows the intersection points between the running median of photometric errors from Fig. \ref{Fig:lim_mag} and the error limits of 0.05, 0.1 and 0.15 mags. These points are defined and referred to hereafter as the NCCS median limiting magnitudes. The limiting magnitudes here are slightly brighter than those in \cite{GOR10I}, since the calibration field used in \cite{GOR10I} was exposed at airmass $\sim1$, whereas the most NCCS fields were at airmass $\sim2$.

\begin{table}[ht!]
 \caption{Limiting Magnitudes \label{Tab:lim_mag}}
 \begin{center}
 \begin{tabular}{|c||c|c|c|}
\cline{2-4}
\multicolumn{1}{c||}{~}  &
\multicolumn{3}{c|}{$\Delta$V, $\Delta$R, $\Delta$I [mag]}\\\hline
Band & 0.05 & 0.10 & 0.15\\ \hline\hline
V & 18.55 & 19.51 & 20.34 \\\hline
R & 18.68 & 19.73 & 21.01 \\\hline
I & 17.73 & 18.64 & 19.23 \\\hline
\end{tabular}
\end{center}
\end{table}

\subsection{Photometry}
\label{sec:photo_comp}

\begin{figure*}[ht!]
 \begin{center}
  \includegraphics[angle=0,width=0.4725\textwidth]{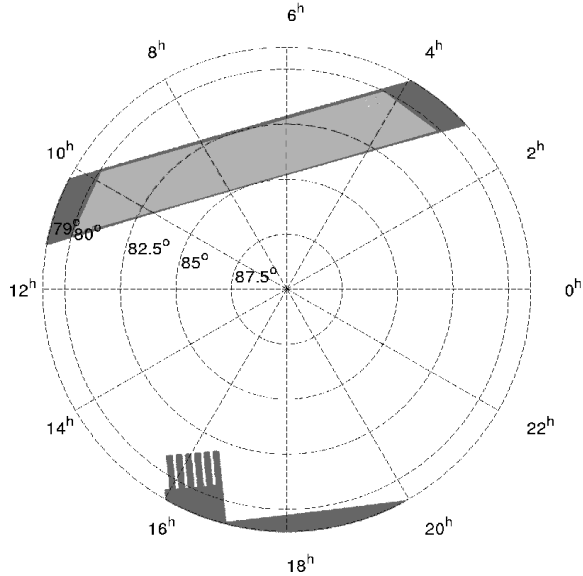} 
\end{center}
 \caption{The dark gray area shows the SDSS total coverage in the North Celestial Pole region. The light gray area shows the common SDSS and NCCS region selected for comparison tests.
    \label{Fig:SDSS}}
 \end{figure*}

We describe below a photometry comparison between the NCCS and the Sloan Digital Sky Survey. The SDSS Data Release 8 (DR8; \citealt{AIH11}) data for the NCCS region were downloaded from the SDSS/SkyServer site\footnote{http://skyserver.sdss3.org/dr8/en/} on May 7, 2012. The SDSS coverage for $\delta\geq79^{\circ}$ is shown in Fig.~\ref{Fig:SDSS} and consists of 768,351 SDSS sources.

The stripe shown in Fig.~\ref{Fig:SDSS} was extracted from the SDSS and the NCCS common areas to ensure identical coverage between the two surveys. SDSS contributed 509,143 sources within the selected stripe, and NCCS - 153,883. 

Among the downloaded SDSS data many sources were separated by less than the NCCS matching criterion of 3 arcsec. These multiple SDSS sources have been observed as single sources in NCCS. One possible reason for this source multiplication is the separation of bright galaxies into several different sources in SDSS (``source shredding''). Another comes from the different pixel scales of two surveys and the different typical seeing. The SDSS sources were re-analyzed to fit the NCCS matching criterion and the weighted mean magnitudes and positions were selected to represent multiple SDSS entries, reducing the total number of SDSS sources to 160,965. There are less NCCS sources within the selected stripe than SDSS sources; this can be explained by the different survey depths. 

\begin{figure*}[ht!]
 \begin{center}
  \includegraphics[angle=0,width=0.97\textwidth]{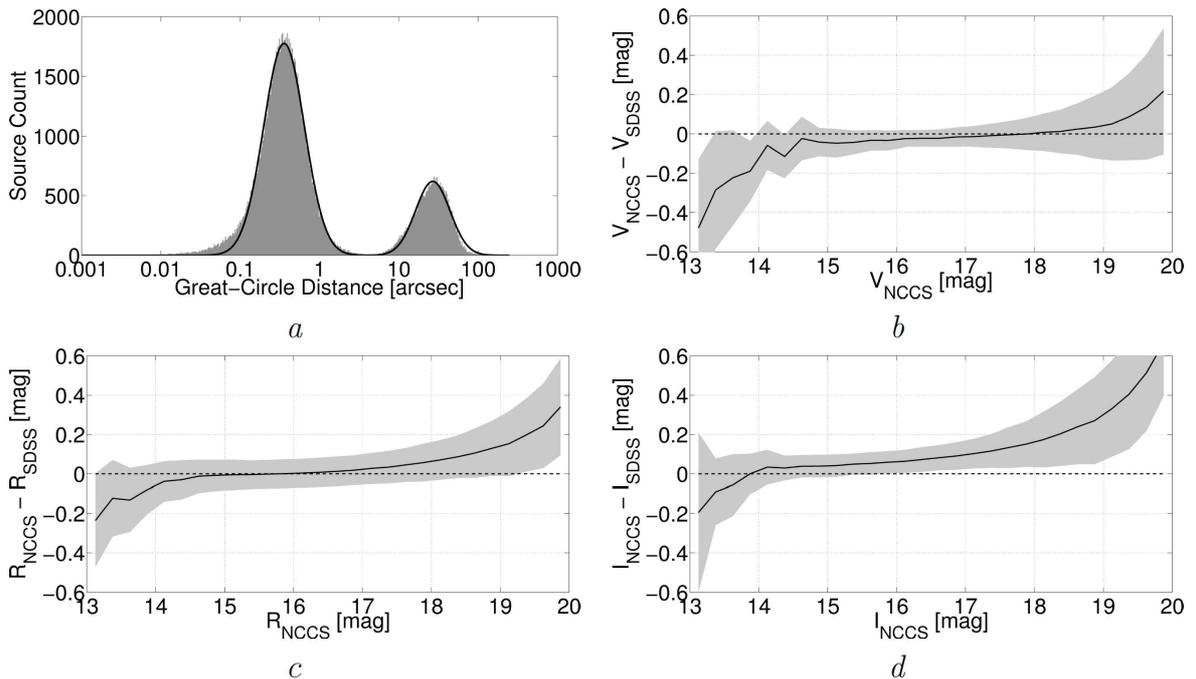} 
 \end{center}
\caption{Panel $a$: Source count as a function of the great-circle distance between the nearest NCCS and SDSS sources. The gray bars are the measured distribution, the black solid line is the fitted two-peak-Gaussian model. Panels $b$, $c$ and $d$: The NCCS photometry compared to the SDSS photometry. The black solid line shows the mean magnitude difference and the gray shaded region marks its standard deviation as a function of V, R and I magnitude respectively, as obtained from a Gaussian fit at different magnitude bins.
    \label{Fig:NCCS_SDSS_photo}}
 \end{figure*}

Each NCCS source within the comparison stripe in Fig.~\ref{Fig:SDSS} was matched to the nearest SDSS source. Fig.~\ref{Fig:NCCS_SDSS_photo}$a$ shows the distribution of great-circle distances between the nearest NCCS and SDSS sources. The distribution was fitted with a two-peak-Gaussian model assuming that the distribution is log-normal. The first peak of the distribution corresponds to the successfully matched sources and contains $\sim78\%$ sources. The second peak corresponds to mismatched sources and contains primarily faint objects. There is a clear separation between the two peaks at $\sim$4.22 arcsec. The peak of the `successful match' is located at $0.36^{+0.29}_{-0.16}$ arcsec, which is sufficiently close to 0 and corresponds to the astrometric accuracy of the NCCS ($<0.3$ arcsec). The `mismatch' peak is at $27^{+17}_{-11}$ arcsec, which corresponds to the mean spatial separation between the NCCS sources within the selected region.

The photometric comparison used only the successfully matched 120,086 sources ($\sim78\%$). The SDSS \textit{ugriz} magnitudes were transformed to the Johnson-Cousins system using the \cite{CHO08} transformations. Fig.~\ref{Fig:NCCS_SDSS_photo}$b,c,d$ shows the comparison between the NCCS and the SDSS photometry. 
The slope of the distributions shown in Fig.~\ref{Fig:NCCS_SDSS_photo}$b,c,d$ for sources with $V,R,I\lesssim14.5$ mag is dominated by the SDSS saturation (for sources with $V,R,I\lesssim11.5$ mag the NCCS optical data are also saturated), while the change for $g$, $r$, $i\gtrsim19.5$ mag can be attributed to a Malmquist bias (see \citealt{CHO08}).

\begin{figure*}[ht!]
 \begin{center}
 \includegraphics[angle=0,width=0.97\textwidth]{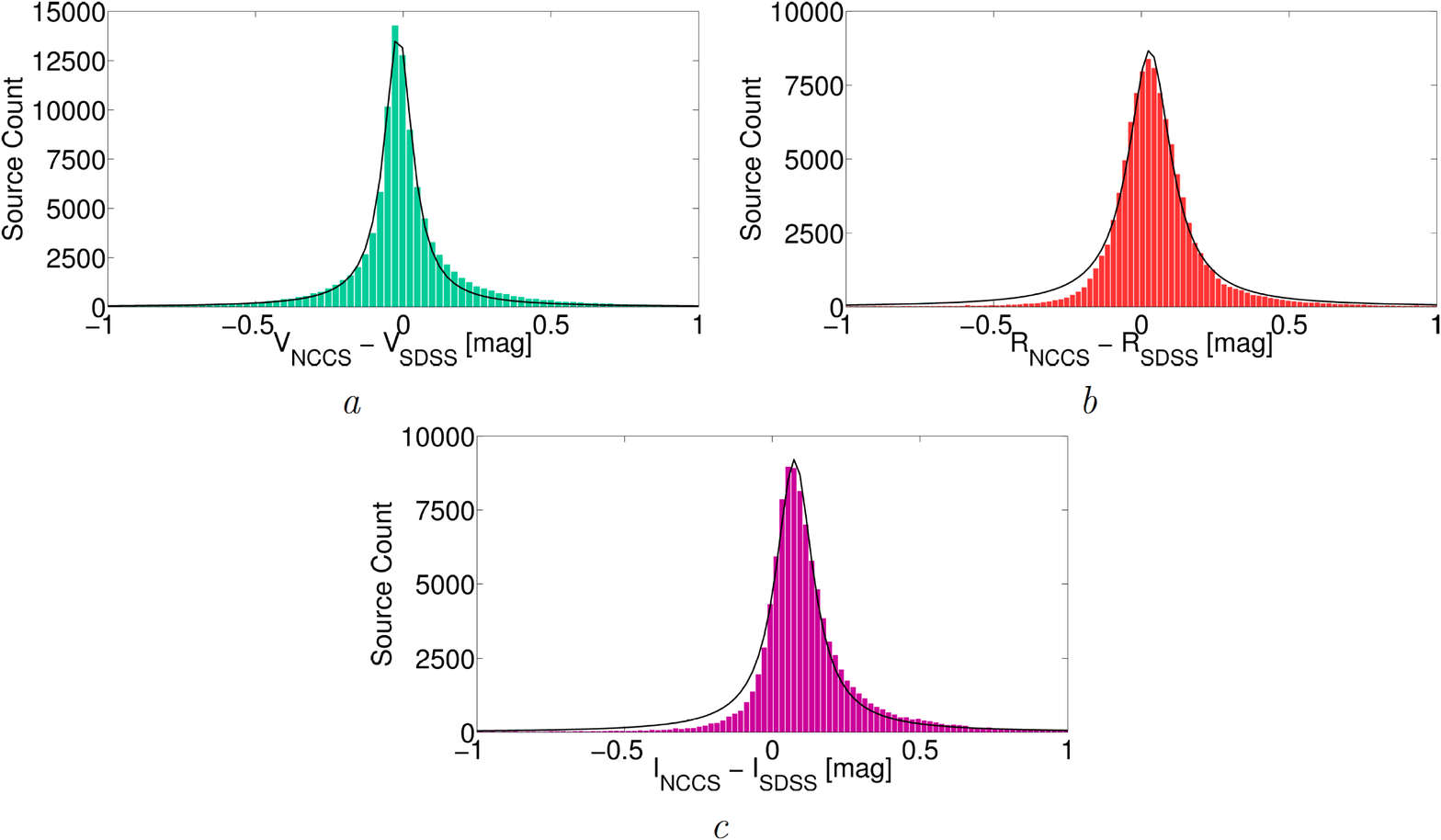} 
 \end{center}
 \caption{The distribution of photometric differences between NCCS and SDSS fitted with a Lorentzian model (black solid line). Panels \textit{a}, \textit{b} and \textit{c} show the data for the V, R and I band respectively.
    \label{Fig:dVdRdI}}
 \end{figure*}

Fig.~\ref{Fig:dVdRdI} shows the distributions of photometric differences between NCCS and SDSS for sources fainter than the SDSS saturation limit but brighter than the NCCS limiting magnitude ($14.5<V<19.51$, $14.5<R<19.73$ and $14.5<I<18.64$). The distributions were fitted with Gaussian and Lorentzian models and the results are presented in Table \ref{Tab:Gaussian_Lorentzian}. Both models yield similar parameters for all three optical bands, but the Lorentzian model provides a slightly better fit (see the GoFT R$^2$\footnote{The coefficient of determination R$^2$ is defined as R$^2 = 1-\frac{\sum{(y_i-f_i)^2}}{\sum{(y_i-\bar{y})^2}}$, where $y_i$ is the measured data, $\bar{y}$ is the mean value of the data and $f_i$ is the fitted model. R$^2$ is used as a GoFT for non-weighted fits and its values can range from -$\infty$ to 1. In the case that R$^2\ll1$, the measured data is fitted by its average value much better than by the proposed model (see \citealt{MEN09}, Chapter 12).\label{footnote_one}} value in Table \ref{Tab:Gaussian_Lorentzian}). Note that the zero photometric difference between NCCS and SDSS is within the error range for all three optical bands for both Gaussian and Lorentzian models, implying that the NCCS photometry matches that of SDSS.

\begin{table}[ht!]
 \caption{The best-fit parameters obtained for the distributions in Fig.\ref{Fig:dVdRdI}.
\label{Tab:Gaussian_Lorentzian}}
 \begin{center}
 \begin{tabular}{|c||c|c|c||c|c|c|}
\cline{2-7}
\multicolumn{1}{c||}{~}  &
\multicolumn{6}{c|}{Model}\\
\cline{2-7}
\multicolumn{1}{c||}{~} & \multicolumn{3}{c||}{Gaussian} & \multicolumn{3}{c|}{Lorentzian}\\\hline
Band & $\mu$ -mean & $\sigma$ - st.dev. & $R^2$ & $x_0$ - median &
$\gamma$ - HWHM & $R^2$\\\hline\hline
V & -0.015 & 0.062 & 0.941 & -0.017 & 0.058 & 0.992\\\hline
R & 0.029 & 0.094 & 0.989 & 0.028 & 0.087 & 0.990\\\hline
I & 0.078 & 0.082 & 0.968 & 0.075 & 0.076 & 0.986\\\hline
\end{tabular}
\end{center}
\end{table}

\subsection{Completeness and Purity}

\begin{figure*}[ht!]
 \begin{center}
\includegraphics[angle=0,width=0.97\textwidth]{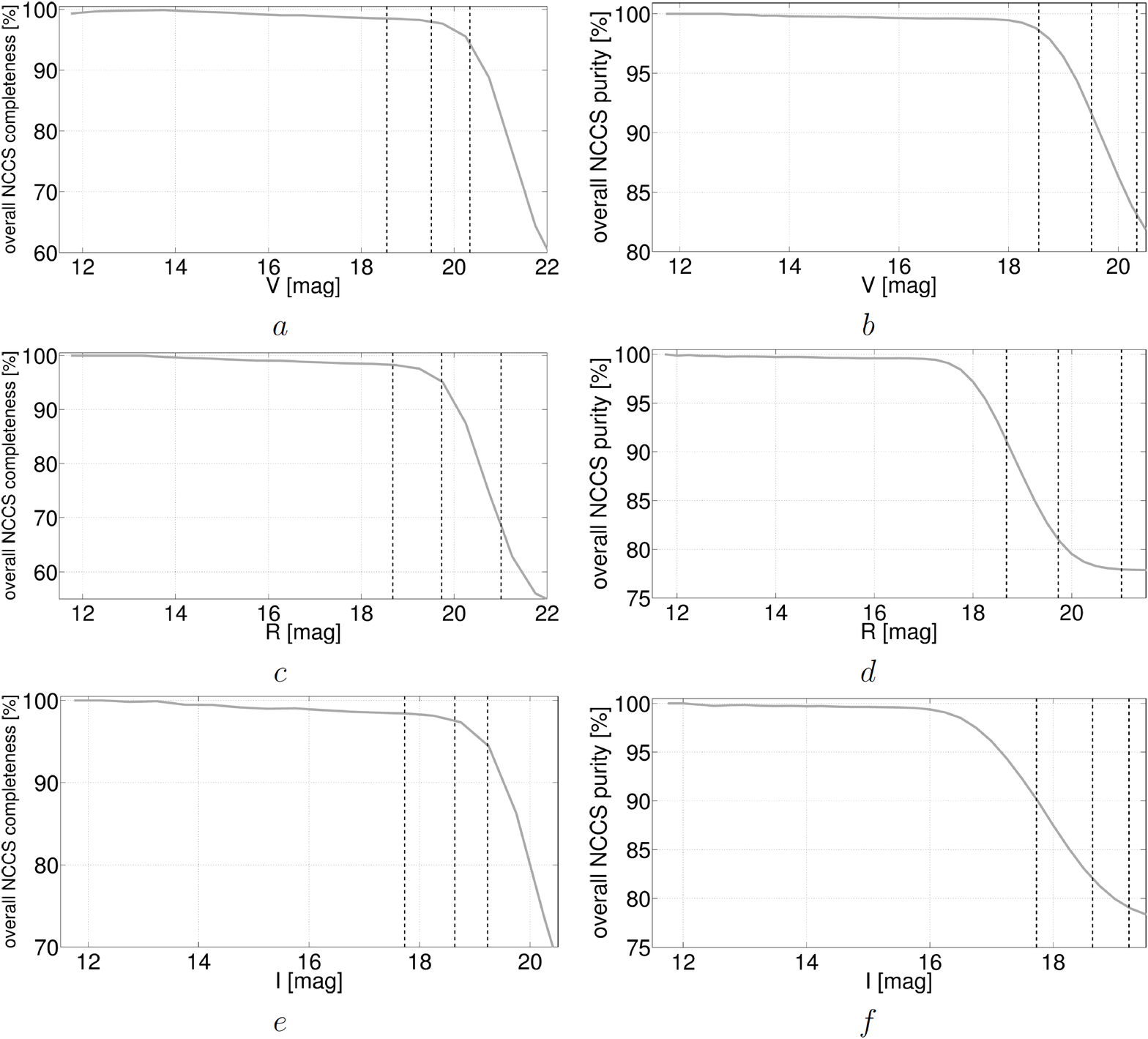} 
 \end{center}
 \caption{Panels $a$, $c$, $e$: overall NCCS completeness for the V, R and I band respectively. Panels \textit{b}, \textit{d}, \textit{f}: overall NCCS purity for the V, R and I band respectively. The black dashed lines show the NCCS limiting magnitudes defined in Table \ref{Tab:lim_mag}.
    \label{Fig:completeness}}
 \end{figure*}

For the completeness test we used only 160,965 rematched SDSS and 153,883 NCCS sources in the stripe plotted in Fig.~\ref{Fig:SDSS}. For each magnitude $m$ we calculate the overall NCCS completeness as a fraction of SDSS sources with $11^m.5 < V,R,I \leq m$ detected by NCCS among the entire SDSS sample of objects with $11^m.5 < V,R,I \leq m$. The completeness vs. magnitude plots are shown for each band in Fig.~\ref{Fig:completeness}$a$,$c$,$e$. 

As expected, the NCCS completeness is close to 100\% in the bright magnitude range, while for faint magnitudes it declines significantly. Table \ref{tab:purity} shows the NCCS overall completeness in the median limiting magnitudes defined in Table \ref{Tab:lim_mag}. We conclude that within the limiting magnitudes corresponding to a median magnitude error of $\Delta$m = 0.1 mag the NCCS catalog is complete to better than 95\% relative to the SDSS. 

\begin{table}[ht!]
 \caption{The overall NCCS completeness and purity in \% in the limiting magnitudes defined in Table \ref{Tab:lim_mag}.\label{tab:purity}}
 \begin{center}
 \begin{tabular}{|c|c||c|c|c|}
\cline{2-5}
\multicolumn{1}{c|}{} & Band & \multicolumn{3}{c|}{limiting magnitude corresponding to}\\\cline{3-5}
\multicolumn{1}{c|}{} & & $\Delta$m = 0.05 mag & $\Delta$m = 0.1 mag & $\Delta$m = 0.15 mag \\\hline\hline
completeness & V & 98.5\% & 98.0\% & 94.3\% \\
purity & & 98.6\% & 91.6\% & 83.0\% \\\hline
completeness & R & 98.3\% & 95.2\% & 68.5\% \\
purity & & 91.2\% & 81.0\% & 77.9\% \\\hline
completeness & I & 98.4\% & 97.5\% & 94.6\% \\
purity & & 90.2\% & 82.0\% & 79.1\% \\\hline
\end{tabular}
\end{center}
\end{table}

The purity of the NCCS optical data was also tested using the SDSS DR8 data. For each magnitude $m$ we calculate the overall NCCS purity as a fraction of NCCS sources with $11^m.5 < V,R,I \leq m$ detected by SDSS among the entire NCCS sample of objects with $11^m.5 < V,R,I \leq m$. The results are shown for each band in Fig.~\ref{Fig:completeness}$b$,$d$,$f$. As expected, the NCCS purity is close to 100\% for bright objects, while for the faint objects the purity decreases significantly for all the bands.

The overall NCCS purity in the limiting magnitudes is shown in Table \ref{tab:purity}. The overall purity of the NCCS catalog in the limiting magnitudes is $\sim$80-90\%, when compared with the SDSS data. Note that the R and I band data are slightly less pure than the V band data, particularly in the faint magnitude range. 

\subsection{Astrometry}
\label{sec:astrometry}

\begin{figure*}[ht!]
 \begin{center}
  \includegraphics[angle=0,width=0.97\textwidth]{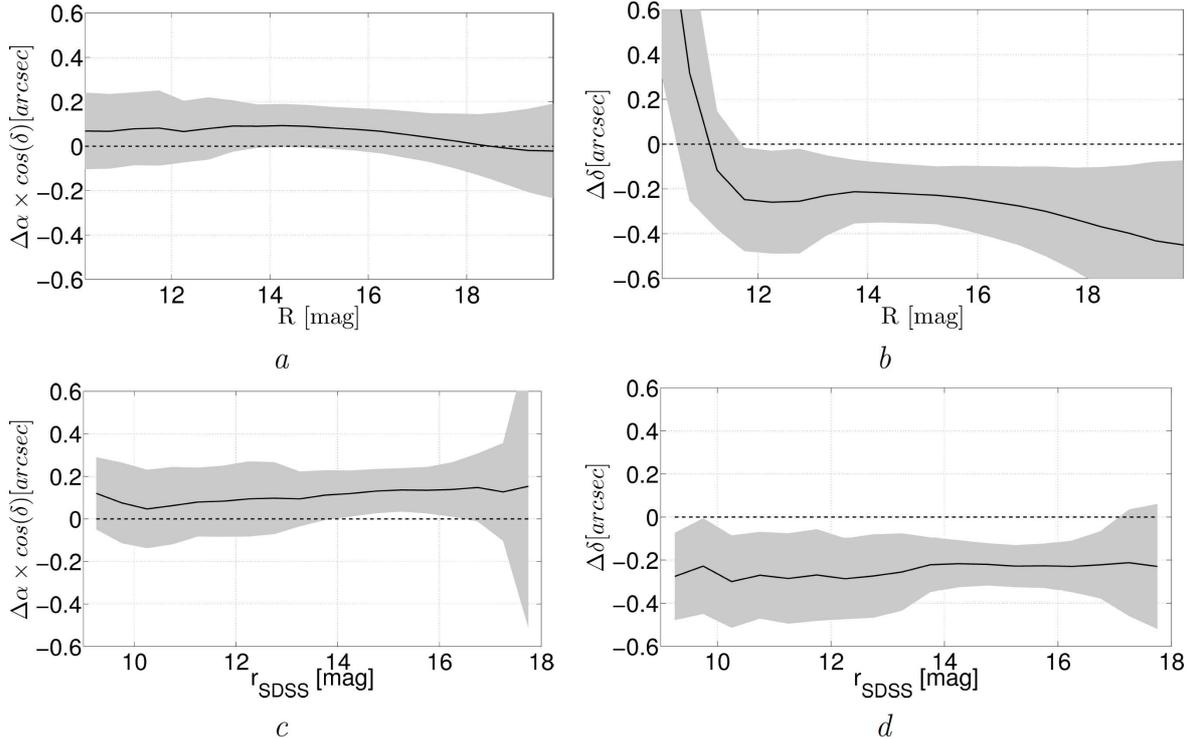}
  \end{center}
 \caption{Astrometry comparison: NCCS vs. SDSS ($a$ and $b$), UCAC3 vs. SDSS ($c$ and $d$). Panels \textit{a} and \textit{b} show the difference in $\alpha$ and $\delta$ respectively, as a function of R, for each individual source. Panels \textit{c} and \textit{d} show the difference in $\alpha$ and $\delta$ respectively, as a function of SDSS \textit{r} magnitude.The black solid line shows the mean positional difference and the gray shaded area shows the standard deviation as obtained from a Gaussian fit to the different magnitude bins.
    \label{Fig:alpha_delta}}
 \end{figure*}

The `successfully matched' 120,086 sources were selected for the astrometry test with the results shown in Fig.~\ref{Fig:alpha_delta}$a,b$. The slope in the declination difference for sources brighter than R = 11.5 mag is explained by NCCS saturation. Note that while the $\alpha$ difference remains almost constant and at the zero level within the error range for the sources R $>11.5$ mag, the $\delta$ difference is $\sim0.3$ arcsec for the entire magnitude range tested. This systematic offset cannot be attributed to the astrometric accuracy of the NCCS, which is $\lesssim$0.2 arcsec, since the NCCS astrometric error is a statistical error increasing the spread but not producing a systematic shift.

Further investigations of the $\delta$ deviation required a similar test with UCAC3 (\citealt{ZAC10}), used here as a source of astrometric standards, against the SDSS sources. Each SDSS unique source was matched to the closest UCAC3 source. Fig.~\ref{Fig:alpha_delta}$c,d$ shows the differences in $\alpha$ and $\delta$ between the two surveys. The differences in both coordinates for $r\gtrsim17$ mag are noisy, consistent with the UCAC3 limiting magnitude R $\approx16.3$ mag (See UCAC3 README file: http://www.nofs.navy.mil/data/fchpix/ucac3\_readme.html and also \citealt{ZAC10}).

The $\alpha$ difference for $r\lesssim17$ mag is almost constant at $\sim$0.1 arcsec, with the zero difference value within the error range (see Fig. \ref{Fig:alpha_delta}\textit{c}) and is explained by the astrometric accuracy of the two surveys: $\lesssim$0.1 arcsec for SDSS (\citealt{AIH11}, see also \citealt{PIE03}) and from 15 to 100 mas for UCAC3 (\citealt{ZAC10}). However, the $\delta$ difference is $\sim$0.3 arcsec for almost the entire magnitude range $r\lesssim17$ mag; this systematic offset cannot be attributed to the reported astrometric accuracy of the two surveys.

This issue was noted by \cite{AHI11e} who compared between the astrometry of SDSS DR7 (\citealt{SCH10}) and SDSS DR8. They found a systematic shift between two astrometric solutions in regions near the North Celestial Pole. The source for the astrometric calibrations of both SDSS DR7 and SDSS DR8 was UCAC2 (\citealt{ZAC04}). However, UCAC2 covers the northern hemisphere only up to $\delta\cong+41^\circ$. For astrometric calibrations at high declinations the SDSS DR7 used ``\textit{...a supplemental set of UCAC results in an internal USNO product known as `r14'}'' (\citealt{AHI11e}), while SDSS DR8 used USNO-B (\citealt{MON03}) catalogue. \cite{AHI11e} found that ``\textit{in the regions not covered by UCAC2 (starting northward of roughly 41$^\circ$ declination), the DR8 astrometry is offset in the mean 240 mas to the north and 50 mas to the west relative to the r14 catalogue.}'' These offsets are similar to those found here between SDSS DR8 and NCCS, thus we conclude that the systematic deviation was imprinted in the SDSS DR8 data, which was used for the present comparison.

\subsection{PESS - Point-Extended Source Separation}
\label{ch:pess}

The goal of the PESS procedure presented here is to set distinct borders on the relative FWHM in each optical band so as to reconstruct with maximal similarity the SDSS star/galaxy classification (for the SDSS star/galaxy classification algorithm see:\\ https://www.sdss3.org/dr8/algorithms/classify.php). The PESS procedure presented here is different from that described in \cite{GOR10I}. The catalogue FWHM was normalized by the mode FWHM for all the objects in the same field. This, since the seeing can vary significantly between the nights and also during the same night. Moreover, the majority of the objects in the survey area are stars. The mode value of FWHM is, therefore, the typical stellar FWHM and we are interested only in the relative FWHM value to define whether the object is likely to be a star or a galaxy. Finally, the mean FWHM can be calculated from relative values for an object exposed several times. For objects imaged multiple times the mean value of normalized FWHM called here $\epsilon$FWHM was selected.

\begin{figure*}[ht!]
 \begin{center}
\includegraphics[angle=0,width=0.97\textwidth]{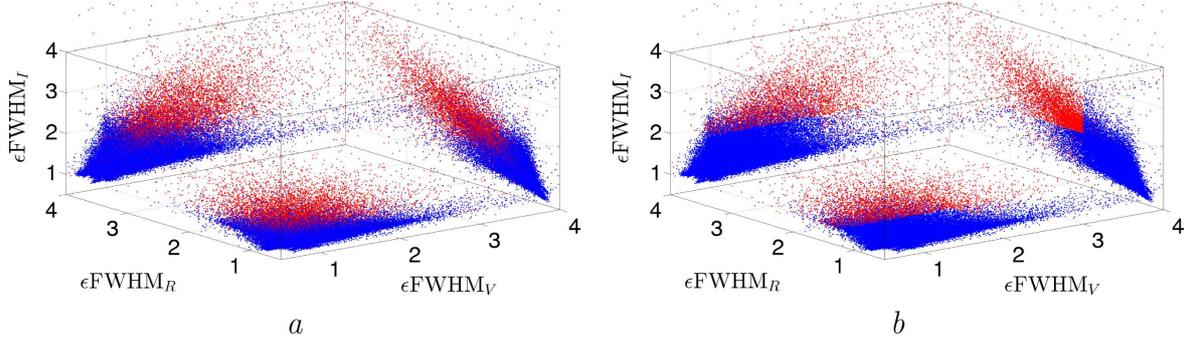} 
 \end{center}
 \caption{2D projections of the 3D distribution of sources' relative FWHM's. (\textit{a}): the original distribution with respect to the SDSS PESS. (\textit{b}): the reconstructed distribution with our own criteria implemented. Blue dots represent point sources, red dots represent extended ones.
    \label{Fig:PESS}}
 \end{figure*}

We select only the 120,086 'successfully matched' NCCS sources and show in Fig.~\ref{Fig:PESS}\textit{a} the 3D distribution of their relative FWHMs with respect to the original SDSS PESS. We used separately the distribution of $\epsilon$FWHMs in each optical band and adopted an arbitrary $\epsilon$FWHM value, defining all objects with $\epsilon$FWHM less than this value as 'stars' and all with $\epsilon$FWHM greater than this value as 'galaxies'. We then checked the number of false definitions relative to SDSS and changed the adopted $\epsilon$FWHM limit to minimize the number of false definitions. We found identical cutoff values for $\Delta$m$<$0.1 mag and $\Delta$m$<$0.15 mag, with $\epsilon$FWHM cutoff being 2.38 (V), 1.82 (R), and 1.79 (I).

We now define a value called `PESS score' for each object; it is equal to 0 if the object is defined as 'star' in all the three optical bands, 1 - if the object is defined as 'galaxy' in one of the bands, 2 - if the object is defined as 'galaxy' in two bands and 3 if the object is defined as 'galaxy' in all the bands. Bin 0 of the PESS score was found to contain almost exclusively SDSS stars, bin 1 contains mostly SDSS stars, bin 2 contains mostly SDSS galaxies and bin 3 contains almost exclusively SDSS galaxies. Therefore, all NCCS sources with a PESS 0 or 1 are defined as 'point-like objects', while those with PESS 2 or 3 are classified as `extended objects', accepting a small fraction of cross-contamination.

The successful PESS rate for all the sources brighter than the limiting magnitudes with $\Delta$m$<$0.1 mag is $\sim$93.6\% relative to the SDSS PESS, and $\sim$92.2\% for those with $\Delta$m$<$0.15 mag. In this manner the SDSS PESS procedure is reconstructed at $\gtrsim$ 92\% accuracy using only object morphology. In Fig.~\ref{Fig:PESS}\textit{b} we show the 3D distribution of relative FWHMs for point and extended sources as defined by our PESS procedure. We conclude that, within the NCCS magnitude limits, our PESS procedure can separate stars and galaxies almost as well as SDSS.

\section{Summary}

We presented here the North Celestial Cap Survey (NCCS), combining our original V, R and I observations with publicly available UV and IR data. This is the first CCD-based optical catalogue for the $\delta\geq80^\circ$ region, covering $\sim$320 deg$^2$ of the sky, being 1-2 mag shallower than SDSS. Only 1,569,292 sources were observed in all three optical bands, while 4,102,998 distinct sources were detected in at least one optical band. Among the final catalogue sources 1,357,281 ($\sim$86.5\%) have WISE counterparts and 301,482 ($\sim$19.2\%) have GALEX counterparts. Only 24,872 objects were detected in all nine optical-UV-IR bands of the survey.

The astrometric accuracy of the optical NCCS data, based on the UCAC3 catalogue, is $\lesssim$0.2 arcsec and the photometry is good to $\sim$0.15 mag for sources brighter than V = 20.3, R = 21.0 and I = 19.2 mag. The NCCS astrometry and photometry were tested against and were found to be consistent with the SDSS data. We also reproduced the SDSS point-extended source separation with $>$92\% efficiency.

\section{Acknowledgements}

This study makes use of the data from USNO CCD Astrograph Catalogue 3, which is a product of the United States Naval Observatory.

Funding for the SDSS and SDSS-II has been provided by the Alfred P. Sloan Foundation, the Participating Institutions, the National Science Foundation, the U.S. Department of Energy, the National Aeronautics and Space Administration, the Japanese Monbukagakusho, the Max Planck Society, and the Higher Education Funding Council for England. The SDSS Web Site is http://www.sdss.org/.

The SDSS is managed by the Astrophysical Research Consortium for the Participating Institutions. The Participating Institutions are the American Museum of Natural History, Astrophysical Institute Potsdam, University of Basel, University of Cambridge, Case Western Reserve University, University of Chicago, Drexel University, Fermilab, the Institute for Advanced Study, the Japan Participation Group, Johns Hopkins University, the Joint Institute for Nuclear Astrophysics, the Kavli Institute for Particle Astrophysics and Cosmology, the Korean Scientist Group, the Chinese Academy of Sciences (LAMOST), Los Alamos National Laboratory, the Max-Planck-Institute for Astronomy (MPIA), the Max-Planck-Institute for Astrophysics (MPA), New Mexico State University, Ohio State University, University of Pittsburgh, University of Portsmouth, Princeton University, the United States Naval Observatory, and the University of Washington.

This publication makes use of data products from the Wide-field Infrared Survey Explorer, which is a joint project of the University of California, Los Angeles, and the Jet Propulsion Laboratory/California Institute of Technology, funded by the National Aeronautics and Space Administration.

Based on observations made with the NASA Galaxy Evolution Explorer. GALEX is operated for NASA by the California Institute of Technology under NASA contract NAS5-98034.

This project makes use of MATLAB software (MATLAB 6.1 or later, The MathWorks Inc., Natick, MA, 2000).

We would like to express our gratitude to the technical and administrative staff of the Wise Observatory for allocating time to our project and for their priceless assistance in observatory hardware and software troubleshooting.

We would like to thank the anonymous referee, whose remarks greatly improved this paper.

\singlespacing

\newpage
\begin{center}
  {\bf APPENDIX A: Example of the Final Optical Catalogue Data.} 
	\end{center}
	The catalogue is stored in file \textit{catalog\_final.mat}. Column contents and explanations are given in Table\ref{tab:catalog_final}.

\begin{table*}[ht!]
\begin{center}
\includegraphics[angle=0,width=1.28\textwidth,angle=90]{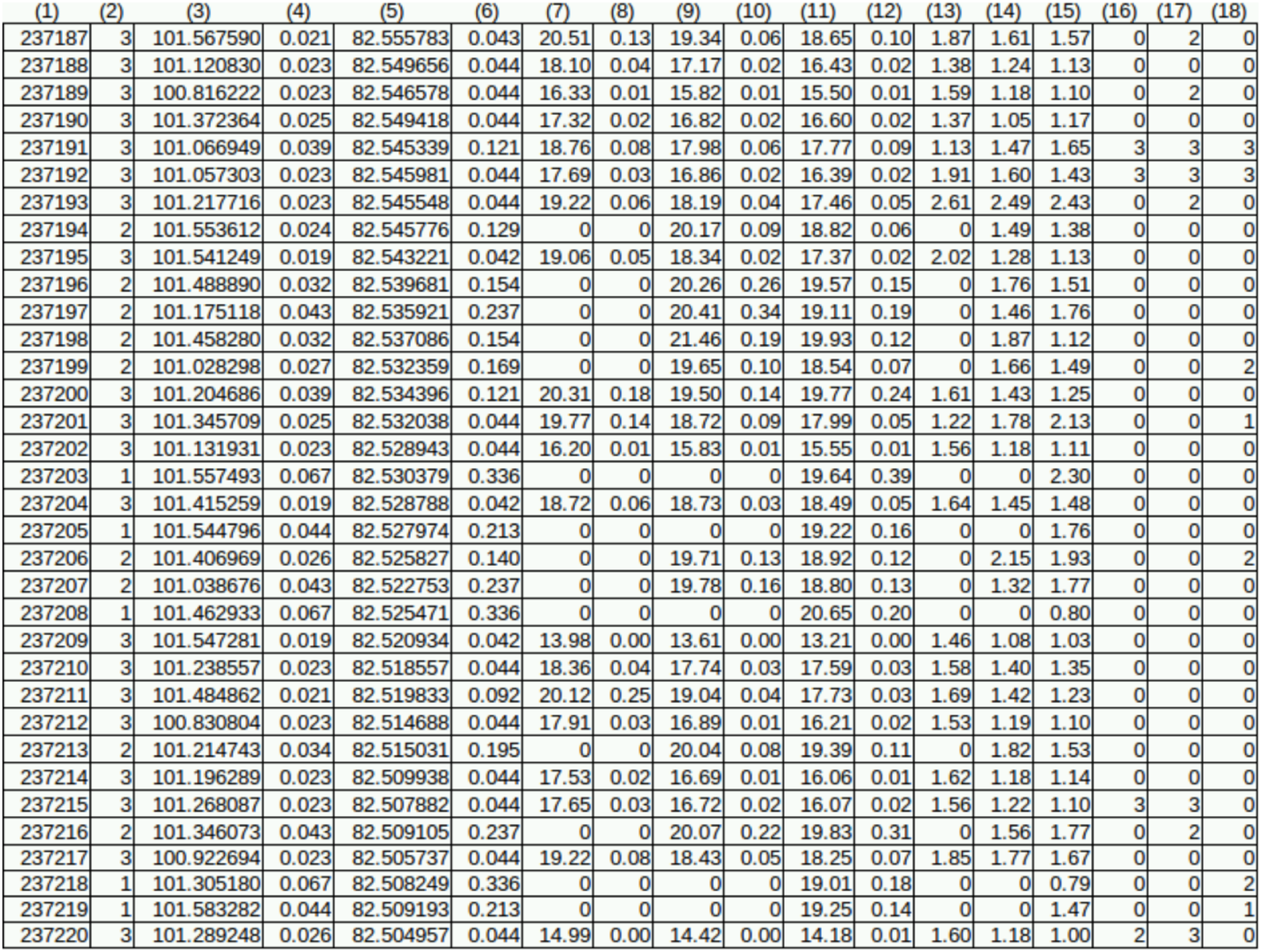}
 \end{center}
\end{table*}

\newpage
\begin{center}
  {\bf APPENDIX B: Example of the Colour-Colour and Colour-Magnitude Diagrams for the NCCS Point-Like Objects.} 
	\end{center}
	The NCCS data used in this example is extracted for the circular region of 1 deg radius centered on $(\alpha,\delta)=(12.108333^\circ, +85.255^\circ)$. These coordinates correspond to the center of the open cluster NGC 188. The colourbar shows the number of NCCS stars per magnitude bin.

\begin{table*}[ht!]
 \begin{center}
\begin{tabular}{c}
 \includegraphics[angle=0,width=0.97\textwidth]{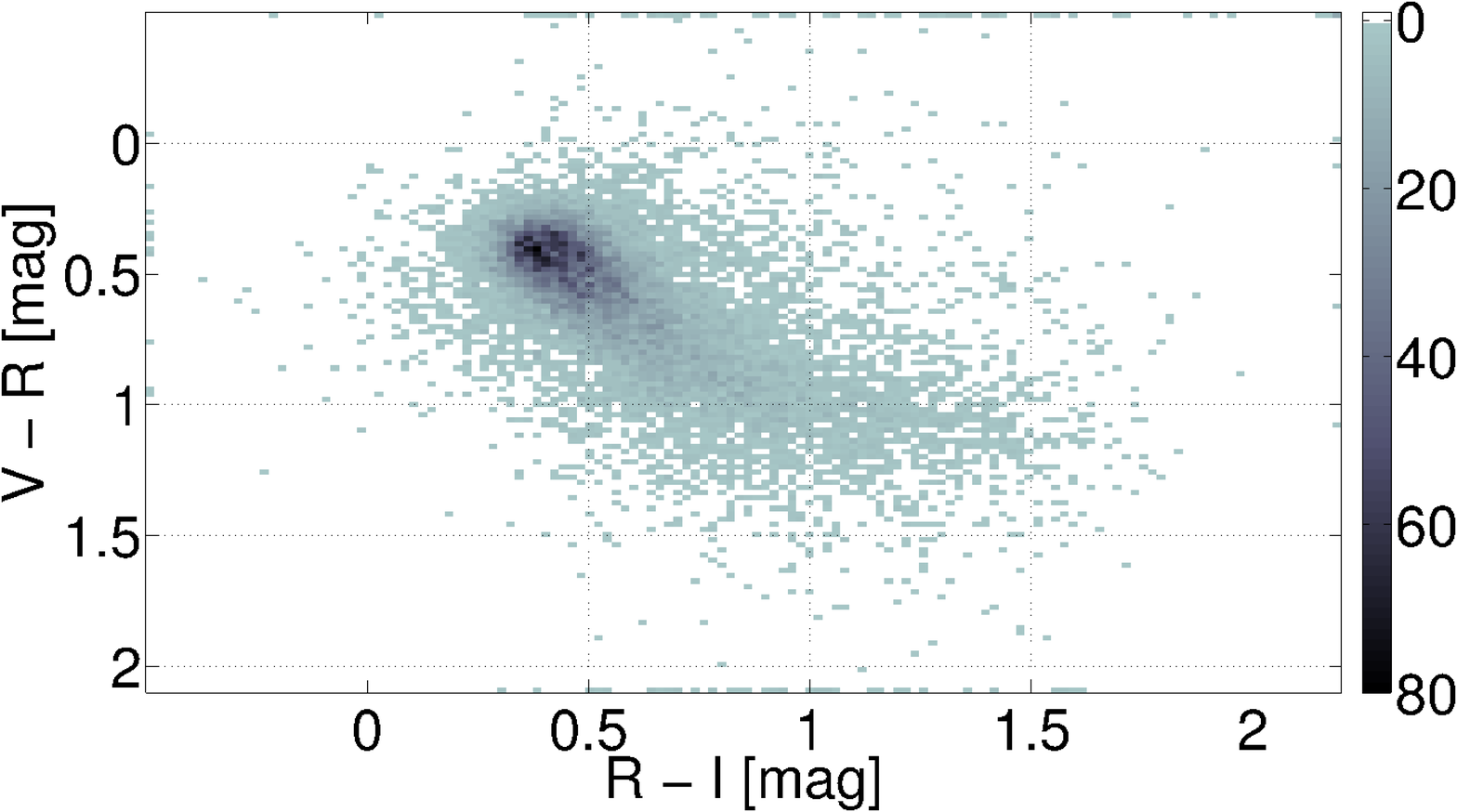}\\
 \\
 \includegraphics[angle=0,width=0.97\textwidth]{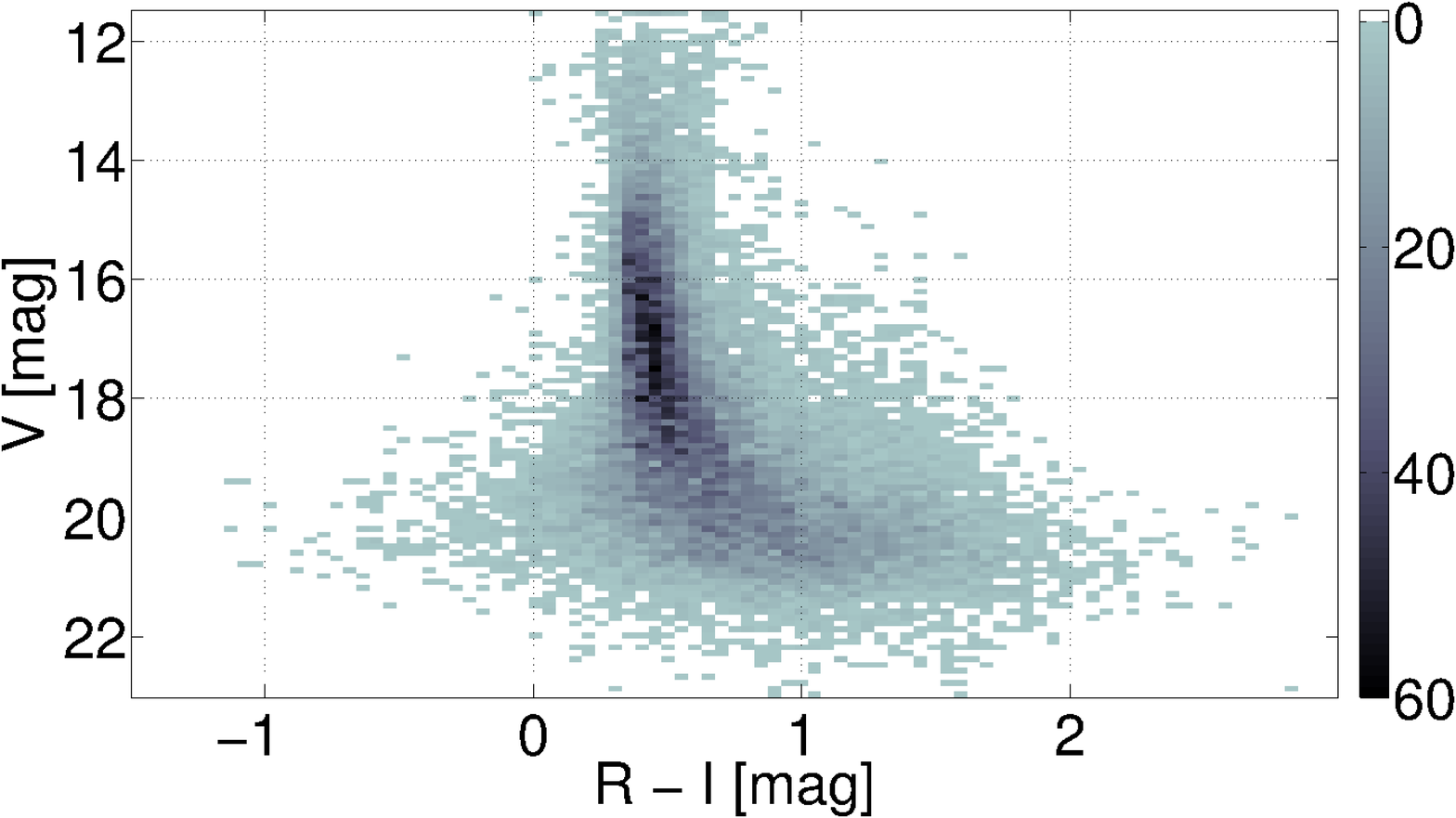}\\
 \end{tabular}
\end{center}
 \end{table*}

\newpage
\begin{center}
  {\bf APPENDIX C: Example of the Final Merged Catalogue Data.} 
	\end{center}
	The catalogue is stored in file \textit{catalog\_clean\_wise\_galex.mat}. Column contents and explanations are given in Table \ref{tab:catalog_cwg}.
\begin{table*}[ht!]
 \begin{center}
\begin{tabular}{cc}
 \includegraphics[angle=0,width=1.27\textwidth,angle=90]{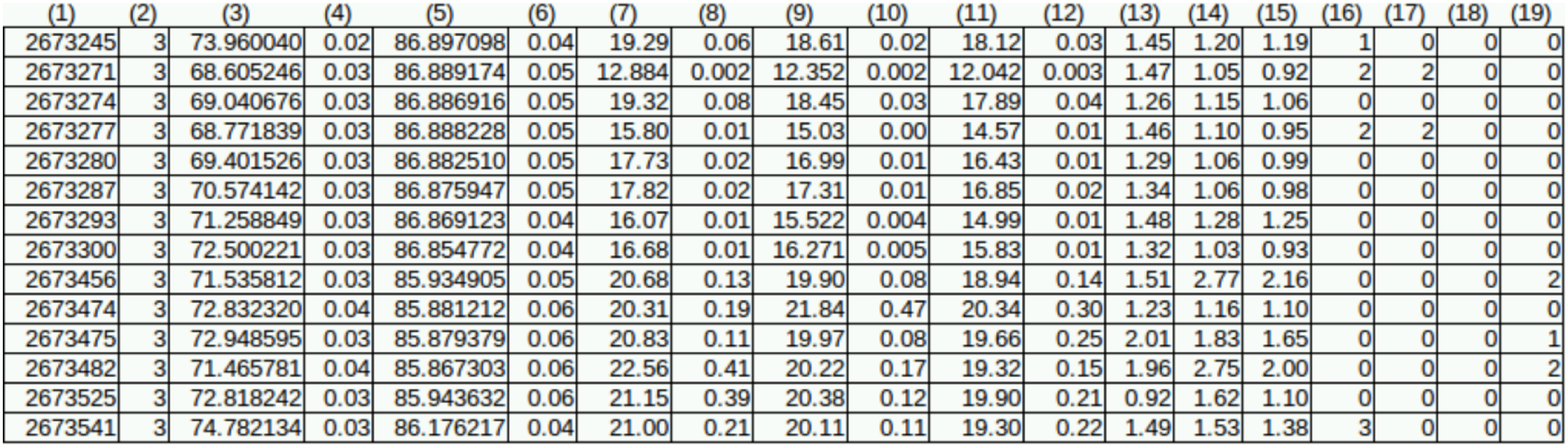} &
 \includegraphics[angle=0,width=1.27\textwidth,angle=90]{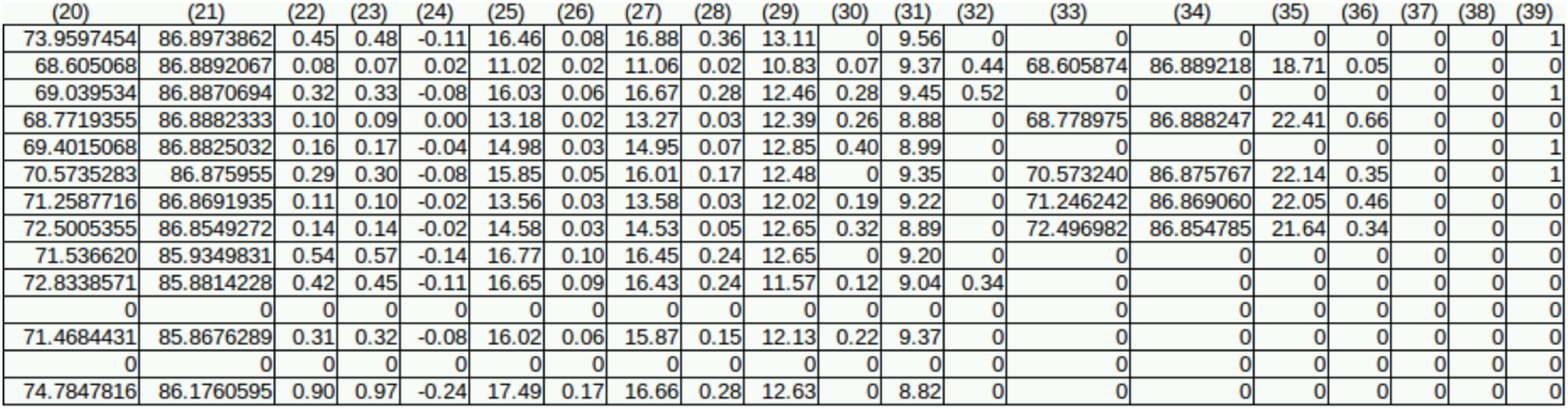}
 \end{tabular}
\end{center}
 \end{table*}


\begin{thebibliography}{}

\bibitem[Ahn et al.(2012)]{SDSS_DR9} Ahn, C.~P., Alexandroff, R., Allende Prieto, C., et al.\ 2012, \apjs, 203, 21

\bibitem[Aihara et al.(2011)]{AIH11} Aihara, H., Allende Prieto, C., An, D., et al.\ 2011, \apjs, 193, 29

\bibitem[Aihara et al.(2011 - Erratum)]{AHI11e} Aihara, H., Allende Prieto, C., An, D., et al.\ 2011, \apjs, 195, 26

\bibitem[Almoznino(2007)]{ALM07} Almoznino, E.\ 2007, Bulletin of the Astronomical Society of India, 35, 209

\bibitem[Barlow(1989)]{BAR89} Barlow, R.\ 1989, \textit{Statistics. A guide to the use of statistical methods in the physical sciences}, The Manchester Physics Series, New York: Wiley, 1989,

\bibitem[Bertin \& Arnouts(1996)]{BER96} Bertin, E., \& Arnouts, S.\ 1996, \aaps, 117, 393

\bibitem[Bovy et al.(2012)]{BOV12} Bovy, J., Rix, H.-W., Liu, C., et al.\ 2012, \apj, 753, 148

\bibitem[Brosch(1992)]{BRO92} Brosch, N.\ 1992, \qjras, 33, 27 

\bibitem[Brosch et al.(2008)]{BRO08} Brosch, N., Polishook, D., Shporer, A., et al.\ 2008, \apss, 314, 163

\bibitem[Chonis \& Gaskell(2008)]{CHO08} Chonis, T.~S., \& Gaskell, C.~M.\ 2008, \aj, 135, 264

\bibitem[Cousins(1976)]{COU76} Cousins, A.~W.~J.\ 1976, \memras, 81, 25

\bibitem[Dobashi et al.(2005)]{DOB05} Dobashi, K., Uehara, H., Kandori, R., et al.\ 2005, \pasj, 57, 1 

\bibitem[Gorbikov \& Brosch(2010)]{GOR10a} Gorbikov, E., \& Brosch, N.\ 2010, \mnras, 401, 231

\bibitem[Gorbikov et al.(2010)]{GOR10I} Gorbikov, E., Brosch, N., \& Afonso, C.\ 2010, \apss, 326, 203

\bibitem[Gorbikov \& Brosch(2011)]{GOR11} Gorbikov, E., \& Brosch, N.\ 2011, \apss, 335, 217

\bibitem[Johnson(1963)]{JOH63} Johnson, H.~L.\ 1963, Photometric Systems, p. 204 in \textit{Basic Astronomical Data: Stars and Stellar Systems}, ed. K. A. Strand, Chicago, IL: Univ. of Chicago Press, 1963

\bibitem[Juri{\'c} et al.(2008)]{JUR08} Juri{\'c}, M., Ivezi{\'c}, {\v Z}., Brooks, A., et al.\ 2008, \apj, 673, 864 

\bibitem[Kron(1980)]{KRO80} Kron, R.~G.\ 1980, \apjs, 43, 305

\bibitem[Landolt(2009)]{LAN09} Landolt, A.~U.\ 2009, \aj, 137, 4186

\bibitem[Magnier(2007)]{3pi_survey} Magnier, E.\ 2007, The Future of Photometric, Spectrophotometric and Polarimetric Standardization, 364, 153

\bibitem[Mendenhall et al.(2009)]{MEN09} Mendenhall, W., Beaver, R.~J., Beaver, B.~M.\ 2009, \textit{Introduction to probability and statistics}, 13th ed., Belmont, CA : Brooks/Cole, Cengage Learning, 2009

\bibitem[Monet et al.(2003)]{MON03} Monet, D.~G., Levine, S.~E., Canzian, B., et al.\ 2003, \aj, 125, 984

\bibitem[Morrissey et al.(2007)]{MOR07} Morrissey, P., et al.\ 2007, \apjs, 173, 682

\bibitem[Nelder and Mead(1965)]{NEL65} Nelder, J.~A., \& Mead, R. 1965,\newblock \textit{A simplex method for function minimization},\newblock The Computer Journal, 7\penalty0 (4):\penalty0 308--313

\bibitem[Pier et al.(2003)]{PIE03} Pier, J.~R., Munn, J.~A., Hindsley, R.~B., et al.\ 2003, \aj, 125, 1559

\bibitem[Schneider et al.(2010)]{SCH10} Schneider, D.~P., Richards, G.~T., Hall, P.~B., et al.\ 2010, \aj, 139, 2360

\bibitem[Skrutskie et al.(2006)]{SKR06} Skrutskie, M.~F., et al.\ 2006, \aj, 131, 1163 

\bibitem[Tody(1986)]{TOD86} Tody, D.\ 1986, \procspie, 627, 733

\bibitem[Wright et al.(2010)]{WRI10} Wright, E.~L., Eisenhardt, P.~R.~M., Mainzer, A.~K., et al.\ 2010, \aj, 140, 1868

\bibitem[Zacharias et al.(2004)]{ZAC04} Zacharias, N., Urban, S.~E., Zacharias, M.~I., et al.\ 2004, \aj, 127, 3043

\bibitem[Zacharias et al.(2010)]{ZAC10} Zacharias, N., Finch, C., Girard, T., et al.\ 2010, \aj, 139, 2184

\end{thebibliography}
\end{document}